\documentclass[aps,prc,twocolumn,groupedaddress,showpacs,showkeys,
nofootinbib,fleqn]{revtex4}
\usepackage{epsfig,amsmath,amssymb,graphicx,tabularx,dcolumn,bm,longtable}
\usepackage{verbatim}  

\begin{document}

\title{A New Efficient Method for Hartree-Fock-Bogoliubov Calculations of Weakly Bound Nuclei}

\author{M.~Stoitsov}
\affiliation{Department of Physics, Graduate School of Science, Kyoto University, Kyoto 606-8502, Japan}
\affiliation{Department of Physics and Astronomy, University of Tennessee, Knoxville, Tennessee 37996, USA}
\affiliation{Physics Division,  Oak Ridge National Laboratory, P.O. Box 2008, Oak Ridge,Tennessee 37831, USA}
\affiliation{Institute of Nuclear Research and Nuclear Energy, Bulgarian Academy of Sciences, Sofia-1784, Bulgaria}

\author{N.~Michel}
\affiliation{Department of Physics, Graduate School of Science, Kyoto University, Kyoto 606-8502, Japan}

\author{K.~Matsuyanagi}
\affiliation{Department of Physics, Graduate School of Science, Kyoto University, Kyoto 606-8502, Japan}

\date{\today}

\begin{abstract}
We propose a new method to solve the Hartree-Fock-Bogoliubov equations for weakly bound nuclei, which works for both spherical and axially deformed cases. 
In this approach,
the quasiparticle wave functions are expanded in a complete set of analytical P\"{o}schl-Teller-Ginocchio and Bessel/Coulomb wave functions. 
Correct asymptotic properties of the quasiparticle wave functions are endowed in the proposed algorithm. Good agreement is obtained with the results of the
Hartree-Fock-Bogoliubov calculation using box boundary condition for a set of benchmark spherical and deformed nuclei.
\end{abstract}

\pacs{21.10.-k,21.30.+y,21.60.Jz}

\maketitle

\section{Introduction}
\label{Intro}
The study of nuclei far from stability is an increasingly important part of contemporary nuclear physics. 
This topic is related to newly created radioactive beams facilities,
allowing more experiments on nuclei beyond the stability line.
The new experimental opportunities on nuclei with extreme isospin ratio and weak binding bring new phenomena 
which inevitably require a universal theoretical description of nuclear properties for all nuclei. 
The current approach to the problem is the nuclear density functional theory which implicitly rely on Hartree-Fock-Bogoliubov (HFB) theory, 
unique in its ability to span the whole nuclear chart.

The HFB equations can be solved in coordinate space using box boundary condition \cite{Bul80, Dob84}.
This approach (abbreviated HFB/Box in this paper) has been used as a standard tool in the description of spherical nuclei \cite{Dob96}. 
Its implementation to systems with deformed equilibrium shapes is much more difficult, however. Different
approaches have been developed to deal with this problem,
such as the two-basis method \cite{Gal94, Ter97a, Yam01}, the canonical-basis framework \cite{Rei97, Taj98, Taj04} and
basis-spline techniques in coordinate-space calculations
developed for axially symmetric nuclei \cite{Ter03, Obe03}.
These algorithms are precise, but time-consuming.

Configuration-space HFB diagonalization is a useful alternative to coordinate-space calculations  
whereby the HFB solution is expanded in a complete set of single-particle states.
In this context, the harmonic oscillator (HO) basis
turned out to be particularly useful.
Over the years, many configuration-space HFB codes using the HO basis (abbreviated HFB/HO) have been developed,
employing either the Skyrme or the Gogny effective interactions \cite {Gog75, Gir83, Egi80, Egi95, Dob04, hfbtho}, 
or using a relativistic Lagrangian \cite{Rin96} in the context of the relativistic Hartree-Bogoliubov theory.
In the absence of fast coordinate-space methods to obtain
deformed HFB solutions, the configuration-space approach has proved to be a very fast and efficient alternative
allowing large-scale calculations \cite{gshpt02,hfbtho}.

Close to drip lines, however, the continuum states start playing an increasingly important role and it becomes necessary 
to treat the interplay of both continuum and deformation effects in an appropriate manner. 
Unfortunately, none of the existing configuration-space HFB techniques manage to incorporate continuum effects.

The goal of the present work is to find an efficient numerical scheme to solve HFB equations for spherical and axially deformed nuclei, 
which properly takes the continuum effects into account. We will denote this problem as continuum HFB (CHFB). 
Aiming at treating spherical and deformed nuclei on the same footing, we rely on the configuration-space HFB approach.

The HO basis has important numerical advantages;
for example, the use of the Gauss-Hermite quadrature allows fast evaluation of matrix elements. 
On the other hand, its Gaussian asymptotics prevents from expanding
systems with large spatial extension, such as halo nuclear states. This problem can be successfully fixed by using the transformed HO basis (THO) \cite{sto04}. 
The latter transforms the unphysical Gaussian fall-off of HO states into a more physical exponential decay. 
Neither HO nor THO bases, however, are able to provide proper discretization of the quasiparticle continuum.
This has repercussions already at the HFB level, for which the HO and THO bases cannot reproduce simultaneously all asymptotic properties  
of nuclear densities (see Sec.~\ref{results_section}). 
While this shortcoming is obvious for the HO basis, 
it also arises for the THO basis because
the latter can provide only one type of asymptotic form, 
i.e.~the one inserted in the scaling function 
defining the THO wave functions \cite{hfbtho}.
Hence, the THO basis fails to reproduce asymptotic properties, as asymptotic behavior is 
different for respective channels: proton and neutron, normal and pairing densities, different angles for the deformed case. 
In fact, differences between calculations using the THO and the 
coordinate-space bases have been noticed in pairing properties of nuclei 
(see Sec.~\ref{results_section} and Ref.~\cite{MarioUmar}). This indicates that THO calculations may not always be fully accurate even 
in the nuclear region and necessitate careful check of obtained results. 
For the aim of carrying out quasiparticle random phase approximation (QRPA) 
calculations with the HFB quasiparticle representation, 
the HO and THO bases are very likely to be insufficient as they cannot 
provide accurate quasiparticle wave functions in the continuum region.

Obviously, a more practical basis is needed.
The Gamow Hartree-Fock (GHF) basis \cite{ghf} would be appropriate, as it has been demonstrated 
that it can provide the correct asymptotic of loosely bound nuclear states. However, it implies the use of complex symmetric
matrices. Moreover, the presence of basis states which increase exponentially in modulus leads to numerical divergences, 
unless the costly two-basis method is employed \cite{nicolas}.

As we plan to consider bound HFB ground states only, it is more advantageous numerically to employ Hermitian completeness relations, 
whose radial wave functions are real.
They are either bound, thus integrable, or oscillate with almost constant amplitude, 
so that we are free from the numerical cancellation problems associated with the Gamow states. 
It should be stressed that we can generate a Gamow quasiparticle basis using the HFB potentials thus obtained. 
We can then describe resonant excited states by means of the quasiparticle random phase approximation representing the QRPA matrix elements in terms of the
Gamow quasiparticle basis. This serves as an interesting subject for future investigation.

One could expect that the employment of the spherical Hartree-Fock (HF) potential to generate the real continuum HF (CHF) 
complete basis would solve the problem. Unfortunately, the CHF basis is not numerically stable due to the presence of resonances 
in the vicinity of the real continuum.
The continuum states lying close to a narrow resonance are rapidly changing, so that a very dense continuum discretization 
around this resonance is necessary to accurately represent this energy region. Important numerical cancellations 
would occur as continuum wave functions become very large in amplitude close to narrow resonances.

To overcome this difficulty, we adopt a technique based on the exactly solvable P\"{o}schl-Teller-Ginocchio (PTG) potential \cite{ginocchio}. 
The spherical HF potential, seemingly the best candidate to generate a rapidly converging basis expansion, but providing numerically costly
GHF bases or unstable CHF bases, is replaced by a PTG potential fitted to the HF potential if the latter give rise to resonant structure. 
It will be shown that the narrow resonant states of the GHF basis will become bound in the PTG basis, 
so that its scattering states will have no rapid phase shift change, a necessary condition for numerically stable continuum discretization. 
As a result, we obtain a very good basis for HFB calculations. We call this approach HFB/PTG.

To test the feasibility of this new method, we have performed numerical calculations for spherical Ni isotopes near the drip line, $^{84}$Ni~--~$^{90}$Ni, 
for a strongly deformed nucleus $^{110}$Zr, and two HFB solutions for $^{40}$Mg with different, prolate and oblate, deformations. 
Good agreement with THO calculations is obtained.

The paper is organized as it follows. The HFB/PTG algorithm is described in Sec.~\ref{HFB_PTG_scheme_section}, 
while the method used to generate the PTG basis is formulated
in Sec.~\ref{basis_section}. Asymptotic properties of the HFB quasiparticle wave functions are discussed in Sec.~\ref{qp_asymptotic}. 
Results of numerical calculation are presented in Sec.~\ref{results_section}. Brief summary and conclusions are given in Sec.~\ref{conclusion}.
Some technical details related to the PTG basis and calculation of matrix elements are collected in Appendices.

\section{The HFB/PTG approach}
\label{HFB_PTG_scheme_section}

Our aim is to develop an efficient method of solving the CHFB equation
\begin{eqnarray}
\displaystyle
\int d{\bf r'} \sum_{\sigma'}
\left(
\begin{array}{cc}
\displaystyle
      h({\bf r}\sigma, {\bf r'}\sigma') - \lambda &
      \tilde{h}({\bf r}\sigma, {\bf r'}\sigma')   \\ \displaystyle
      \tilde{h}({\bf r}\sigma, {\bf r'}\sigma')
      & -h({\bf r}\sigma, {\bf r'}\sigma') + \lambda
\end{array}
\right)
\nonumber \\
\displaystyle
\times
\left(
\begin{array}{c}
           U(E,{\bf r'}\sigma') \\
           V(E,{\bf r'}\sigma')
\end{array}
\right)
= E
\left(
\begin{array}{c}
           U(E,{\bf r} \sigma) \\
           V(E,{\bf r} \sigma)
\end{array}
\right)
\label{CHFBeq}
\end{eqnarray}
for weakly bound nuclei, which equally works both for spherical and axially 
deformed nuclei. In the above equation, ${\bf r}$ and $\sigma$ are the 
coordinate of the particle in normal and spin space, 
$h({\bf r}\sigma, {\bf r'}\sigma')$ and
$\tilde{h}({\bf r}\sigma, {\bf r'}\sigma')$
denote the particle-hole and the particle-particle (hole-hole) components of 
the single-particle Hamiltonian, respectively,
$U({\bf r} \sigma)$ and $V({\bf r} \sigma)$ the upper and the lower components 
of the single-quasiparticle wave function, and $\lambda$ is the chemical 
potential~\cite{Dob96}. 
For simplicity of notation, the isospin index $q$ is omitted 
in Eq.~(\ref{CHFBeq}), but, of course,  we solve the CHFB equation 
for coupled systems of protons and neutrons.
In this section, we outline the calculational 
scheme and details will be presented in the succeeding sections.

The proposed method to solve the CHFB equations, abbreviated HFB/PTG, consists of the following steps:

(1) One starts with spherical or deformed HFB calculations in the HO basis (HFB/HO). 
This provides a good approximate solution for the HF potential and the effective mass.

(2) One considers a HF potential and an effective mass for each $\ell j$ subspace, 
and fits the associated shifted PTG potential to them when the HF potential possesses bound or narrow resonant states in this $\ell j$ subspace 
(see Sec.~\ref{PTG_fit}). If no such states appear in the HF $\ell j$ spectrum, 
a set of Bessel/Coulomb wave functions \cite{Abr70} is selected for the $\ell j$ partial wave basis.

(3) One diagonalizes the HFB eigenvalue equations in the basis
composed of the PTG and Bessel/Coulomb wave functions. This step continues until self-consistency is achieved.

The use of the Bessel/Coulomb wave functions in step (2)  occurs for partial waves of high angular momentum, 
for which the centrifugal part becomes dominant. As no resonant structure can appear therein in the real HF continuum, 
Bessel/Coulomb wave functions provide a numerically stable set of states for this partial wave.
 For the generation of Coulomb wave functions, one can use the recently published C$++$ code \cite{cwfcomplex} or its FORTRAN alternative  \cite{cwfcomplexF}. 
A complete set of wave functions is thus formed, which will be used as a basis to expand the HFB quasiparticle wave functions.

The necessary truncation of the basis in step (3) implies that spurious effects may eventually appear at very large distances, 
where both the particle density $\rho$ and the pairing density $\tilde{\rho}$ are very small. 
Consequently, quasiparticle wave functions have to be matched to their exact asymptotics at moderate distances 
as it is explained further in Sec.~\ref{qp_asymptotic}. 
In addition, special care must be taken to calculate matrix elements due to the presence of non-integrable scattering states (see Appendix~\ref{ME_appendix}).

When the HF mean-field resulting from the HFB/HO calculation in step (1) is 
deformed, there are several ways to extract the HF potential for each $\ell j$ 
subspace to be used in step (2). Because it is used just as a generator for 
the complete PTG basis, its choice will have little effect on the final HFB 
solution, however. In the present calculation, we therefore adopt a simple 
procedure; the particle-hole part of the HFB/HO potential and the HFB/HO 
effective mass are used in step (2) after averaging their angular and spin 
degrees of freedom. The resulting HF potential is spherical and the same 
for all $\ell j$ subspaces. In such a case, the effect of the spin-orbit 
splitting is not taken into account in the stage of constructing the PTG basis 
but it is of course taken into account in step (3).
This implies to consider a basis generated by a spherical potential, 
which might seem inefficient in the case of large deformation, for which 
deformed bases are more appropriate, as is done with the HO and THO bases.
The deformed nuclei considered in this paper are nevertheless fairly reproduced within this framework (see Sec.~\ref{results_section}). 
If necessary, it is possible to generate a deformed basis 
by diagonalizing the deformed HF potential within the PTG basis,
which can then serve as a particle basis for the HFB problem.

\section{Generation of basis}
\label{basis_section}

\subsection{PTG potentials fitting procedure}
\label{PTG_fit}

The PTG potential has four parameters $\Lambda, s, \nu$ and $a$, 
which have to be determined in each $\ell j$ subspace
(see Appendix ~\ref{PTG_appendix}). For this purpose, we use the spherical 
HF potential and effective mass in a given  $\ell j$ subspace.

\begin{figure}[htb]
\includegraphics[angle=-90,width=0.47\textwidth]{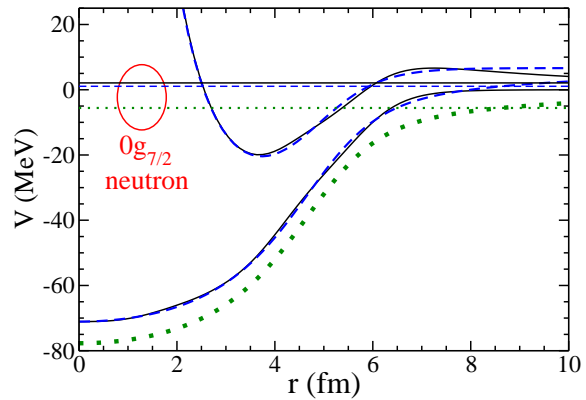}
\protect\caption{\label{shifted_pot} (color online)
The shifted PTG potential, the HF potential 
calculated with the SLy4-force and the unshifted PTG potential
for neutrons in $^{84}$Ni. HF and shited PTG potentials to which centrifugal part is added are provided as well,
and the energies of $0g_{7/2}$ levels for each potential are indicated. 
All data respectively associated to HF, shifted and unshifted PTG potentials
are respectively shown in solid, dashed and dotted lines.}
\end{figure}

The PTG mass parameter $a$ is obtained from the requirement that the PTG 
and the HF effective masses are the same at the origin. 
One first adds the centrifugal term $\displaystyle V_{\ell(\ell+1)} \propto 
\ell(\ell+1)/r^2$ to the nuclear plus Coulomb potential, $V_N + V_C$, 
and determines the height $E_b$ of the centrifugal (plus Coulomb) barrier.
Then, one adds $E_b$ to the PTG potential; 
the resulting potential may be called the shifted PTG potential. 
The parameters $\Lambda$ and $\nu$ are fitted in such a way 
that the $\chi^2$ difference between the shifted PTG potential and 
the HF potential is minimal.
Note that $s$ is directly obtained from $\Lambda$ and $\nu$ values 
during the fit, as it is determined by way of the property that 
the PTG potential of parameters $\Lambda ,s ,\nu $ and $a$
for $r \rightarrow 0$ is equivalent to $s^2$ times the PTG potential of 
parameters $\Lambda, s=1 , \nu$ and $a$.
The reason why we use the barrier height $E_b$ in our fitting procedure
will become apparent by an illustrative example presented below.

To test the fitting procedure and the quality of the resulting PTG basis 
we performed GHF calculations in the coordinate space for the spherical 
nucleus $^{84}$Ni.  Let us examine the quality of single-particle energies 
and wave functions resulting from the shifted PTG potential by comparing 
them with the GHF energies and wave functions for bound and resonance states.

Figure~\ref{shifted_pot} illustrates the PTG fitting procedure and 
compare the results with the GHF ones taking the neutron $0g_{7/2}$ level 
as an example. It is seen that the energy of the bound $0g_{7/2}$ state 
in the original (unshifted) PTG potential (horizontal dotted line) 
become positive after being shifted with $E_b$ (horizontal dashed line) 
and its position agrees in a good approximation with the resonance energy
obtained by the GHF calculation (horizontal solid line). This is due to 
a special feature of the PTG potential, for which the centrifugal potential
decreases exponentially and not as $r^{-2}$ for $r \rightarrow +\infty$ 
(see App.~\ref{PTG_appendix}).
This implies that the centrifugal + shifted PTG potential goes very quickly 
to the constant value, $E_b$, for $r \rightarrow +\infty$.

In this way,
the PTG treatment replaces the GHF resonance with a weakly bound PTG state 
whose wave function will be very similar in the nuclear region. 
Approximating resonant states by weakly bound states in our framework 
resembles the standard two-potential method described in Ref.~\cite{Two_potential}.
Thus, 
one can expect that the fitted PTG potential provides a rapidly converging 
basis for solving the HFB equations.

In fact, it is not necessary to find the PTG potential that exactly minimize 
the $\chi^2$ difference with the HF potential. As the PTG potential is used 
as a basis generator, slight differences with the exact minimum lead only to 
slightly different bases states to expand the HFB quasiparticle wave 
functions, preserving its rapidly converging properties. Thus, one can take 
rather large steps for the $\Lambda, \nu$ variations and few radii for 
the $\chi^2$ evaluation to save computer time, keeping the quality of the basis essentially 
the same.

\subsection{Single-particle energies}
Single-particle energies and widths for neutrons in $^{84}$Ni 
obtained by the GHF calculations are compared with the PTG energies 
in Table~\ref{PTG_neutron_energies}.~~One can clearly see the following facts.

Firstly, the overall agreement between the GHF and the shifted PTG energies 
is good, which means that the PTG potential is flexible enough to reproduce 
the main features of the HF potential.

Secondly, all narrow GHF resonances are represented as weakly bound PTG states 
with upward shifted PTG energies. This is very important because the HFB 
upper (lower) components of quasiparticle states are likely to have large overlaps 
with unoccupied (occupied) weakly bound and narrow resonance states.

We note that the GHF states whose width is larger than 1 MeV, as a rule, are 
not converted to bound PTG states. This is not important, however, because 
scattering states do not exhibit rapid changes in the energy region of broad 
resonances. The broad resonance region can indeed be well represented in terms 
of the continuum basis states.

\begin{table}[ht]
 \centering
 \caption{Neutron GHF levels in $^{84}$Ni calculated with the SLy4 
 Skyrme-force and the surface-type delta pairing interaction  
 (see Sec.~V for the parameters used), which are compared with 
 the PTG estimates. All energies are given in MeV 
 while the width $\Gamma$ is given in keV.}
\label{PTG_neutron_energies}
 \vspace{.2cm}
 \tabcolsep=.2cm
 \begin{ruledtabular}
 \begin{tabular}{cccccc}
\noalign{\smallskip}
   &   \multicolumn{2}{c}{\text{GHF}}           &   \multicolumn{2}{c}{\text{PTG}}     \\
\noalign{\smallskip}
\noalign{\smallskip}
states  &   $\Gamma$ &  $e$ & $e + E_b$ & $e$ \\
\noalign{\smallskip}\hline \\
$0s_{1/2}$ & 0       & -52.38  &  -51.89  &   -51.89   \\
$1s_{1/2}$ & 0       & -24.37  &  -25.55  &   -25.55   \\
$2s_{1/2}$ & 0       & -0.72   &  -0.97   &   -0.97    \\
$0p_{3/2}$ & 0       & -41.25  &  -40.67  &   -41.09   \\
$1p_{3/2}$ & 0       & -12.52  &  -12.95  &   -13.36   \\
$0p_{1/2}$ & 0       & -39.44  &  -38.79  &   -39.22   \\
$1p_{1/2}$ & 0       & -10.67  &  -10.73  &   -11.16   \\
$0d_{5/2}$ & 0       & -29.38  &  -29.50  &   -31.02   \\
$1d_{5/2}$ & 0       & -1.90   &  -1.94   &   -3.46    \\
$0d_{3/2}$ & 0       & -25.20  &  -25.53  &   -27.11   \\
$1d_{3/2}$ & 10.03   &  0.18   &    0.24  &   -1.34    \\
$0f_{7/2}$ & 0       & -17.56  &  -17.45  &   -20.88   \\
$0f_{5/2}$ & 0       & -10.87  &  -12.40  &   -16.01   \\
$0g_{9/2}$ & 0       & -6.11   &  -5.52   &   -11.74   \\
$0g_{7/2}$ & 31.62   &  2.09   &   1.05   &   -5.58    \\
$0h_{11/2}$ & 92.93  &  4.53   &   6.18   &   -3.79    \\
 \noalign{\smallskip}
 \end{tabular}
 \end{ruledtabular}
 \end{table}

\subsection{PTG wave functions}
\begin{figure}[htb]
\includegraphics[angle=-90,width=0.47\textwidth]{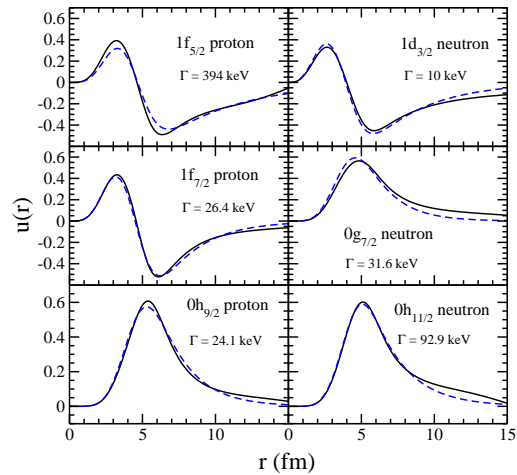}
\protect\caption{\label{fig_overlaps}  (color online)
The PTG  (dashed lines) and  GHF (solid lines) wave functions for various 
resonance states.}
\end{figure}
As illustrated in  Fig.~\ref{fig_overlaps} narrow GHF resonant states bear large overlaps with their associated PTG bound states. 
Hence, the GHF resonant structure present
in the HFB quasiparticle wave functions will be
sustained by the PTG bound states, thus reducing the coupling to the PTG scattering continuum.
\begin{figure}[htb]
\includegraphics[angle=-90,width=0.47\textwidth]{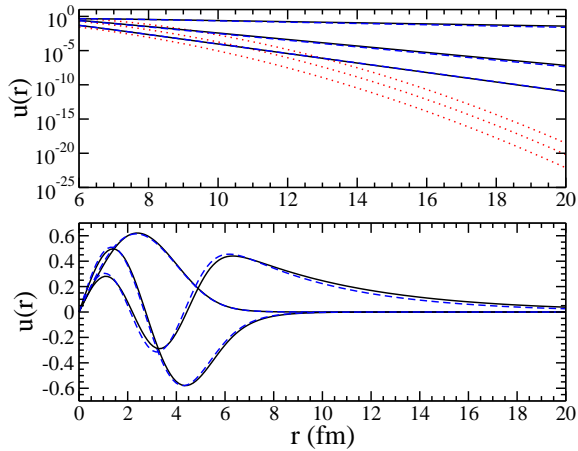}
\protect\caption{\label{s_states_fig}  (color online)
The PTG (dashed lines),  GHF (solid lines) and HO (dotted lines) wave functions including the asymptotic region for the bound $0s_{1/2}, 1s_{1/2}$ and 
$2s_{1/2}$ neutron states 
both in normal scale (lower panel) and logarithmic scale (upper panel).}
\end{figure}

An example indicating the quality of the bound single-particle wave functions resulting from the fitting PTG procedure is shown in Fig.~\ref{s_states_fig} 
for the bound $0s_{1/2}, 1s_{1/2}$ and $2s_{1/2}$ neutron states. In this case, nuclear potential has no centrifugal barrier, so that
the PTG and the HF potentials possess the same asymptotic behavior. Very good agreement between the PTG (dashed lines) 
and the GHF (solid lines) wave functions is thus not surprising. 
The upper panel in Fig.~\ref{s_states_fig} shows the asymptotic region in logarithmic scale where HO wave functions (dotted lines) are also given as a reference. 
Their Gaussian asymptotics cannot reproduce even approximately the exponential decrease of the PTG and GHF wave functions.
\begin{figure}[htb]
\includegraphics[angle=-90,width=0.47\textwidth]{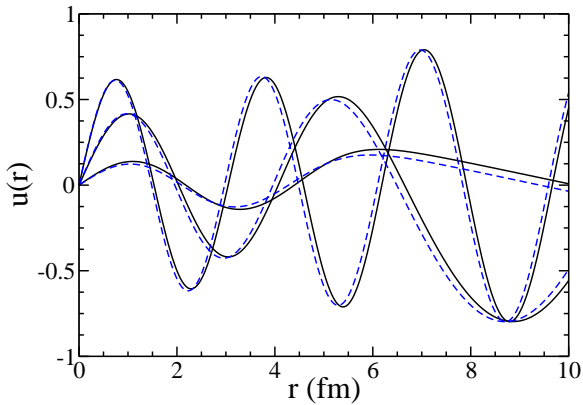}
\protect\caption{\label{s_scat_fig}  (color online)
The PTG  (dashed lines) and  GHF (solid lines) wave functions of the neutron continuum $s$-states
calculated with energies of 0.118 MeV, 9.996 MeV and 66.119 MeV.}
\end{figure}

Neutron continuum $s$-states are illustrated in Fig.~\ref{s_scat_fig}, which are properly reproduced as well
by the scattering states for the PTG potential.
\begin{figure}[htb]
\includegraphics[angle=-90,width=0.47\textwidth]{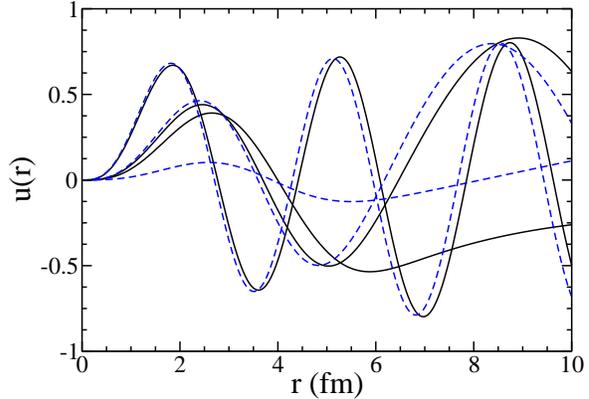}
\protect\caption{\label{d_scat_states}  (color online)
The PTG  (dashed lines) and  the GHF (solid lines) wave functions for the neutron continuum $d_{3/2}$-states
calculated at the same energies as in Fig.~\ref{s_scat_fig}}
\end{figure}
In the cases when a centrifugal (and/or Coulomb) barrier exists,
as illustrated in Fig.~\ref{d_scat_states} for $d_{3/2}$ states, different phase shifts develop in the PTG and GHF continuum states, 
as the PTG potential bears no barrier at large distance.

\section{Quasiparticle wave functions in the asymptotic region}
\label{qp_asymptotic}

The necessary truncation of the basis implies that spurious effects will eventually appear at very large radius,
where both the particle density $\rho$ and the pairing density $\tilde{\rho}$ are very small. 
Consequently, quasiparticle wave functions have to be matched with their exact asymptotics at moderate distance, 
where the asymptotic region has been attained and densities are still large enough for basis expansion to be precise. 
Below we explain how the matching procedure is done for axially deformed nuclei.

In order to deal with the asymptotics of quasiparticle wave functions, we make partial wave decomposition of them:
\begin{equation}
\begin{array}{ccc}
\displaystyle U_{km}({\bf r}\sigma)
= \sum_\alpha U_{km}^{\alpha} \Psi_\alpha ({\bf r})
=\sum_{\ell j} U_{km}^{(\ell j)} (r) \; \mathcal{Y}^{\ell j}_{km}(\Omega), \\~~\\\displaystyle V_{km}({\bf r}\sigma)
= \sum_\alpha V_{km}^{\alpha} \Psi_\alpha ({\bf r})
=\sum_{\ell j} V_{km}^{(\ell j)} (r) \; \mathcal{Y}^{\ell j}_{km}(\Omega), \\
\end{array}
\label{asspartiall}
\end{equation}
where the subscript $k$ specifies the quasiparticle eigenstates together with 
the magnetic quantum number $m$ which is always conserved for both spherical 
and axially symmetric nuclei; 
$\Psi_\alpha ({\bf r})$ are the PTG or Bessel/Coulomb wave functions; $U_{km}^{\alpha}$ and $V_{km}^{\alpha}$ are the HFB expansion coefficients; 
$U_{km}^{(\ell j)} (r)$ and $V_{km}^{(\ell j)} (r)$ are the radial amplitudes with $r = |{\bf r}|$ for the $(\ell j)$ partial wave; 
$\mathcal{Y}^{\ell j}_{m}(\Omega)$ 
denotes a product wave function where the spherical harmonics with the angular 
variables $\Omega$ and the orbital angular momentum $\ell$ is coupled with 
spin to the total angular momentum $j$. 

The partial wave amplitudes, $U_{km}^{(\ell j)}(r)$ and $V_{km}^{(\ell j)}(r)$, 
defined above involve summation over all quantum numbers except the angular momenta $\ell$ and $j$.
In the spherical case, the sums reduce to a single element as $\ell$ and $j$ are good quantum numbers. 
In the asymptotic region, only Coulomb and centrifugal parts remain from
the HFB potentials, so that one can continue the quasiparticle wave functions via their partial wave decompositions and decay constants $k_u$ and $k_v$:
\begin{equation}
\begin{array}{lll}
\displaystyle U_{km}^{(\ell j)} (r) = C_{km}^{(\ell j)+} 
H^{+}_{\ell,\eta_u}(k_u r)
+ C_{km}^{(\ell j)-} H^{-}_{\ell,\eta_u}(k_u r), \\~~\\
\displaystyle V_{km}^{(\ell j)} (r) = D_{km}^{(\ell j)+} 
H^{+}_{\ell,\eta_v}(k_v r), \\~~\\
\displaystyle k_v=\sqrt{ \frac{2m}{\hbar^2} (\lambda - E)},
\displaystyle k_u=\sqrt{ \frac{2m}{\hbar^2} (\lambda + E)},
\end{array}
\label{assc}
\end{equation}
where $E$ denotes the quasiparticle energy, $\lambda$ the chemical potential, $H^\pm_{\ell,\eta}$ the Hankel (or Coulomb) functions, 
$\eta$ being the Sommerfeld parameter, and $C_{km}^{(\ell j)+}$, $C_{km}^{(\ell j)-}$ and $D_{km}^{(\ell j)+}$ are constants to be determined.
Matching is performed using Eq.~(\ref{asspartiall}) at a radius $R_0$ in the asymptotic region where the basis expansion is precise, 
so that $C_{km}^{(\ell j)+}$, $C_{km}^{(\ell j)-}$ and $D_{km}^{(\ell j)+}$ come forward by continuity.
The value of $R_0$ is typically of the order of $10$~fm.

\begin{table*}[ht]
 \centering
 \caption{Results of the HFB/PTG calculation for ground state characteristics  of Ni isotopes close to the neutron drip line, 
which are compared with  results of the HFB/Box calculation.  The SLy4 functional and the surface-type delta pairing\cite{sto04} are used.  
The rms radii are in fm and all other quantities are in MeV.  Proton chemical potential $\lambda_p$ is not provided as pairing correlations
 vanish in the proton space. }
\label{Ni_res_table}
 \vspace{.2cm}
 \tabcolsep=.2cm
 \begin{ruledtabular}
 \begin{tabular}{lrrrrrrrrrrrrrrr}
\noalign{\smallskip}
&  \multicolumn{2}{c}{$^{84}$Ni}  &
&  \multicolumn{2}{c}{$^{86}$Ni}  &
&  \multicolumn{2}{c}{$^{88}$Ni}  &
&  \multicolumn{2}{c}{$^{90}$Ni}  \\
\noalign{\smallskip} &
\multicolumn{1}{c}{\text{HFB/Box}}           &   
\multicolumn{1}{c}{\text{HFB/PTG}} & \smallskip & \multicolumn{1}{c}{\text{HFB/Box}}           &   
\multicolumn{1}{c}{\text{HFB/PTG}} & \smallskip & \multicolumn{1}{c}{\text{HFB/Box}}           &   
\multicolumn{1}{c}{\text{HFB/PTG}} & \smallskip & \multicolumn{1}{c}{\text{HFB/Box}}           &   
\multicolumn{1}{c}{\text{HFB/PTG}}     \\
\hline \\
$\lambda_n$ &   -1.453  &  -1.429  & &   -1.037 &    -1.029  & &  -0.671  &      -0.661 & &   -0.342  &     -0.329   \\
$r_n$ &    4.451 &   4.450  & &    4.528 &     4.526  & &   4.603  &      4.602  & &    4.677  &       4.674  \\
$r_p$ &    3.980  &   3.981  & &    4.001 &     4.001  & &   4.021  &      4.021  & &    4.043  &       4.043  \\
$\Delta_n$ &    1.481  &   1.532  & &    1.667 &     1.658  & &   1.790  &     1.780  & &    1.899  &       1.892  \\
$E^{pair}_n$ &  -30.70   & -30.60   & &  -36.52  &   -35.92   & & -41.98   &    -41.187  & &   -47.158 &     -46.233  \\
$T_n$ & 1084.53   &1085.95   & & 1118.65  &  1118.63   & &1150.71   &   1150.64   & &   1182.52 &    1182.66   \\
$T_p$ &   430.47   & 430.240  & &  425.99  &   426.01   & & 421.71   &    421.72   & &   417.38  &     417.37  \\
$E^{so}_n$ &  -63.379  & -63.177  & &  -61.679 &    61.707  & & -59.558  &     59.681  & &   -56.898 &     -57.889  \\
$E^{Coul}_{dir}$ &  132.94   & 132.90   & &  132.26  &   132.246  & & 131.571  &    131.578  & &    130.947&     130.886  \\
$E^{Coul}_{exc}$ &  -10.138  &  10.136  & &  -10.084 &   -10.085  & & -10.033  &     10.033  & &    -9.980 &      -9.980  \\
$E_{tot}$ & -654.89   &-654.914  & & -656.933 &  -656.877  & &-658.167  &   -658.084  & &   -658.665&    -658.608  \\
 \noalign{\smallskip}
 \end{tabular}
 \end{ruledtabular}
 \end{table*}

\section{Numerical examples}
\label{results_section}

We have made a feasibility test of the HFB/PTG method
for spherical Ni isotopes close to the neutron drip line
and for deformed neutron-rich nuclei $^{110}$Zr and $^{40}$Mg.
All calculations were done using the SLy4 density functional.
For the pairing interaction, we use the surface-type delta pairing 
with the strength $t_0^{'} = -519.9$ MeV fm$^{3}$ for the the 
density-independent part and $t_3^{'} = -37.5t_0^{'}$ MeV fm$^{6}$ for 
the density-dependent part with a sharp energy cut-off at 60 MeV 
in the quasiparticle space. 
They have been fitted to reproduce the neutron pairing gap of $^{120}$Sn. 
These values are consistent with those given in Ref.~\cite{HFBRAD}; 
the slight difference is due to different cut-off procedures, sharp 
cut-off in our case and smooth cut-off in  Ref.~\cite{HFBRAD}.
Below we discuss the major features of the result of calculation. 
We also make a detailed comparison between the HFB/PTG and HFB/Box calculations in the spherical case.

\subsection{Spherical nuclei}

Let us first examine how the result of calculation depends on the truncation of the basis. 
Indeed, the basis has to be truncated at a maximal linear momentum $k_{max}$,
and discretized with $N_{\ell j}$ continuum states per partial wave in the interval $[0:k_{max}]$.
\begin{figure}[htb]
\includegraphics[angle=-90,width=0.47\textwidth]{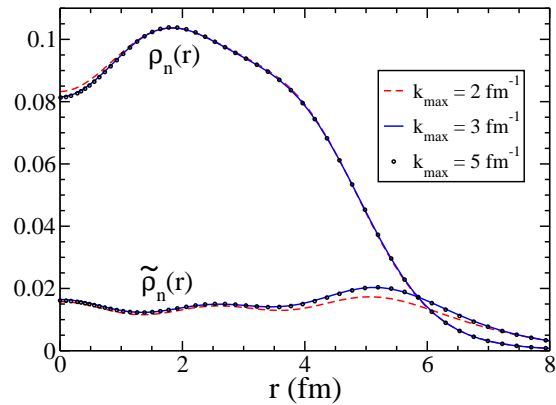}
\protect\caption{\label{rho_dep_kmax}   (color online)
Dependence on $k_{max}$ of the neutron density $\rho_n$
and the neutron pairing density $\tilde{\rho}_n$
calculated for $^{84}$Ni by the HFB/PTG method.}
\end{figure}
Figure~\ref{rho_dep_kmax} shows that the use of values larger than 
$k_{max}=3$ fm$^{-1}$ does not change the results.
Accordingly, in calculations for spherical nuclei, 
we use $k_{max}=5$~fm$^{-1}$ and discretize the continuum with 
$N_{\ell j}=60$ scattering states per partial wave 
(see Ref.~\cite{ghf} for its justification).

\begin{figure*}[htb]
\includegraphics[angle=-90,width=0.80\textwidth]{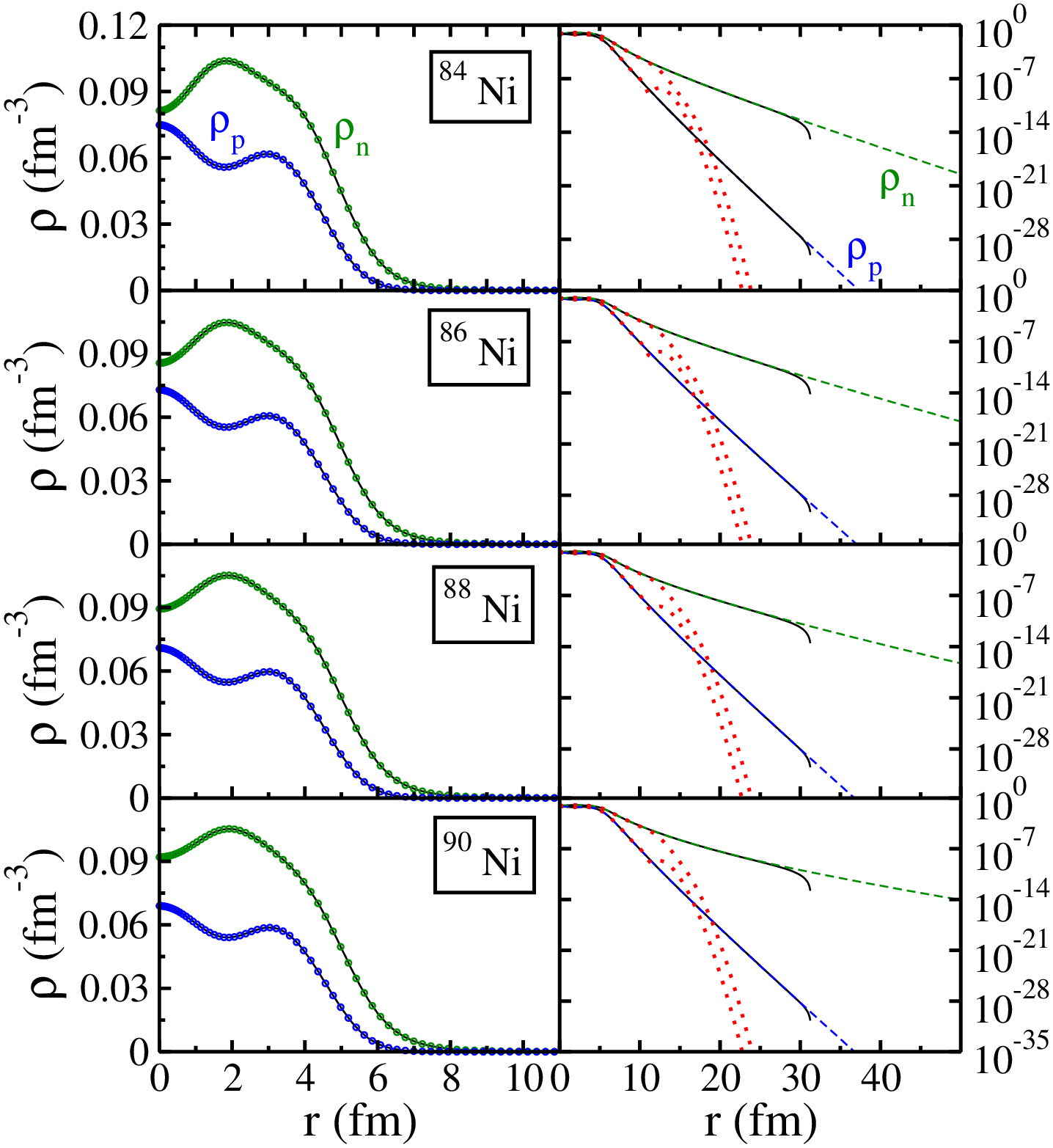}
\protect\caption{\label{fig_spherical_densities}   (color online) The neutron densities $\rho_n$ and proton densities $\rho_p$ 
both in normal (left-hand side) and logarithmic (right-hand side) scales. 
Results of the HFB/Box calculation are displayed by solid lines, 
while those of the HFB/PTG calculations by open circles and dashed lines. 
The HFB/HO densities are also indicated by dotted lines in the right panels for comparison.}
\end{figure*}
\begin{figure*}[htb]
\includegraphics[angle=-90,width=0.80\textwidth]{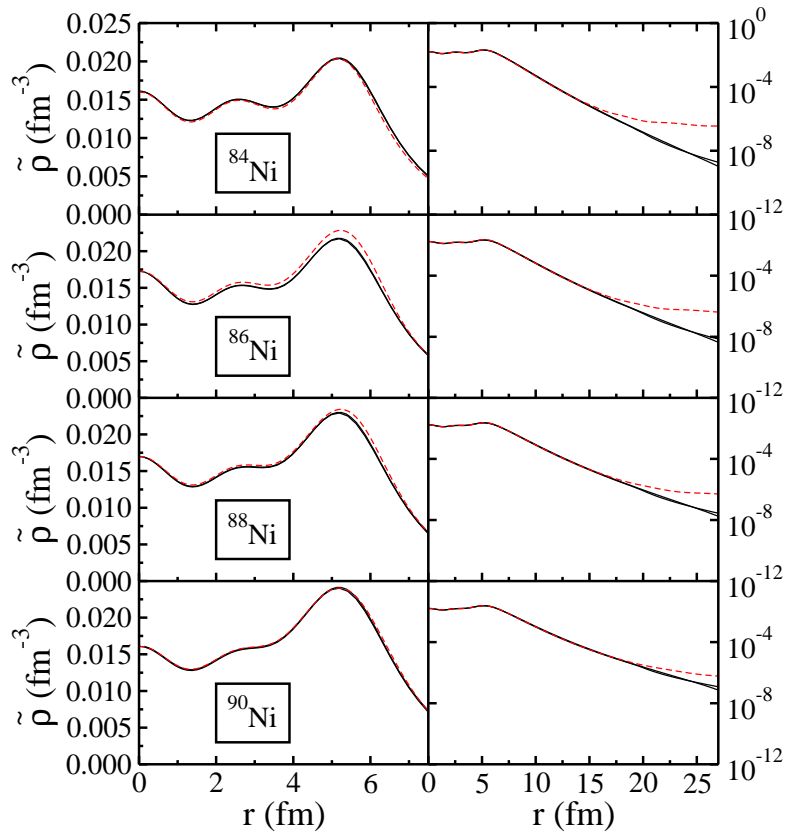}
\protect\caption{\label{fig_spherical_pairing_densities}   (color online) The neutron pairing densities $\tilde{\rho}_n$
in normal (left-hand side) and logarithmic (right-hand side) scales. There are no pairing correlations in the proton channel. 
Results of the HFB/Box and HFB/PTG calculations are displayed both by solid lines, as they are almost indistinguishable, 
while HFB/THO pairing densities are represented by dashed lines.}
\end{figure*}
\begin{figure*}[htb]
\includegraphics[angle=-90,width=0.80\textwidth]{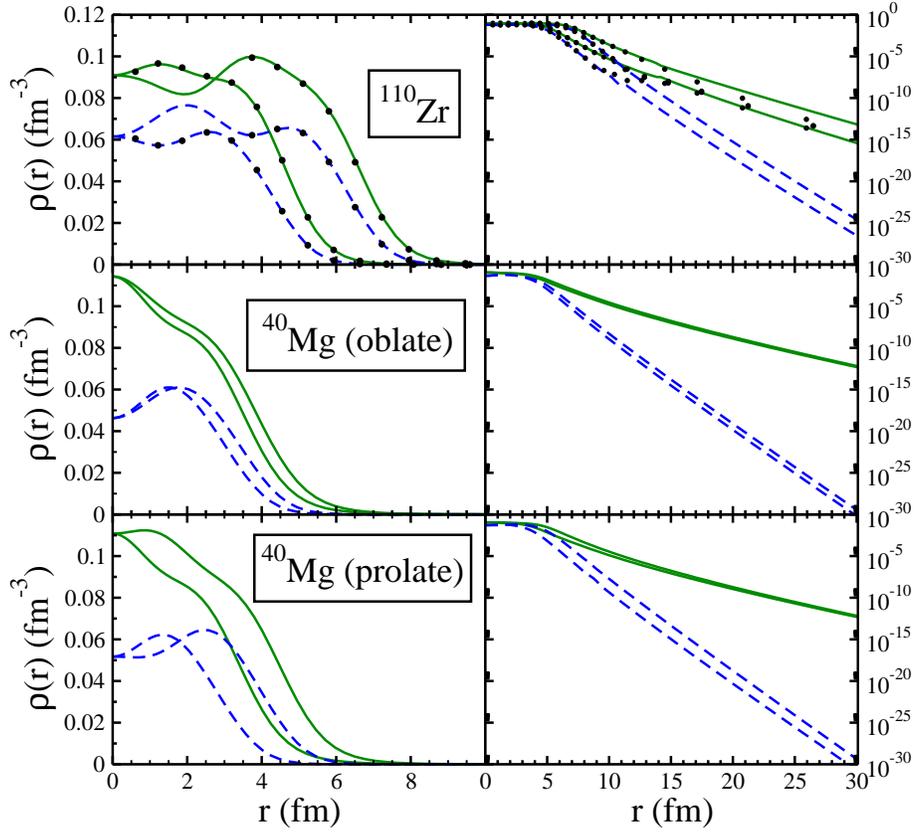}
\protect\caption{\label{fig_deformed_dens}  (color online)
The neutron and proton densities of the prolately deformed nucleus $^{110}$Zr~($\beta = 0.40$), respectively calculated by
the HFB/PTG (respectively solid and dashed lines) and HFB/THO (circles) methods in normal (top left) and logarithmic (top right) scale. 
They are given along the long and short axes of deformation, easily identified from the figure.
The neutron and proton densities of $^{40}$Mg calculated by the HFB/PTG method for two states with different deformations 
(oblate $\beta = -0.09$ and prolate $\beta = 0.26$) 
in normal (middle and bottom left) and logarithmic (middle and bottom right) 
scale are also provided with the same line convention.}
\end{figure*}
\begin{figure*}[htb]
\includegraphics[angle=-90,width=0.80\textwidth]{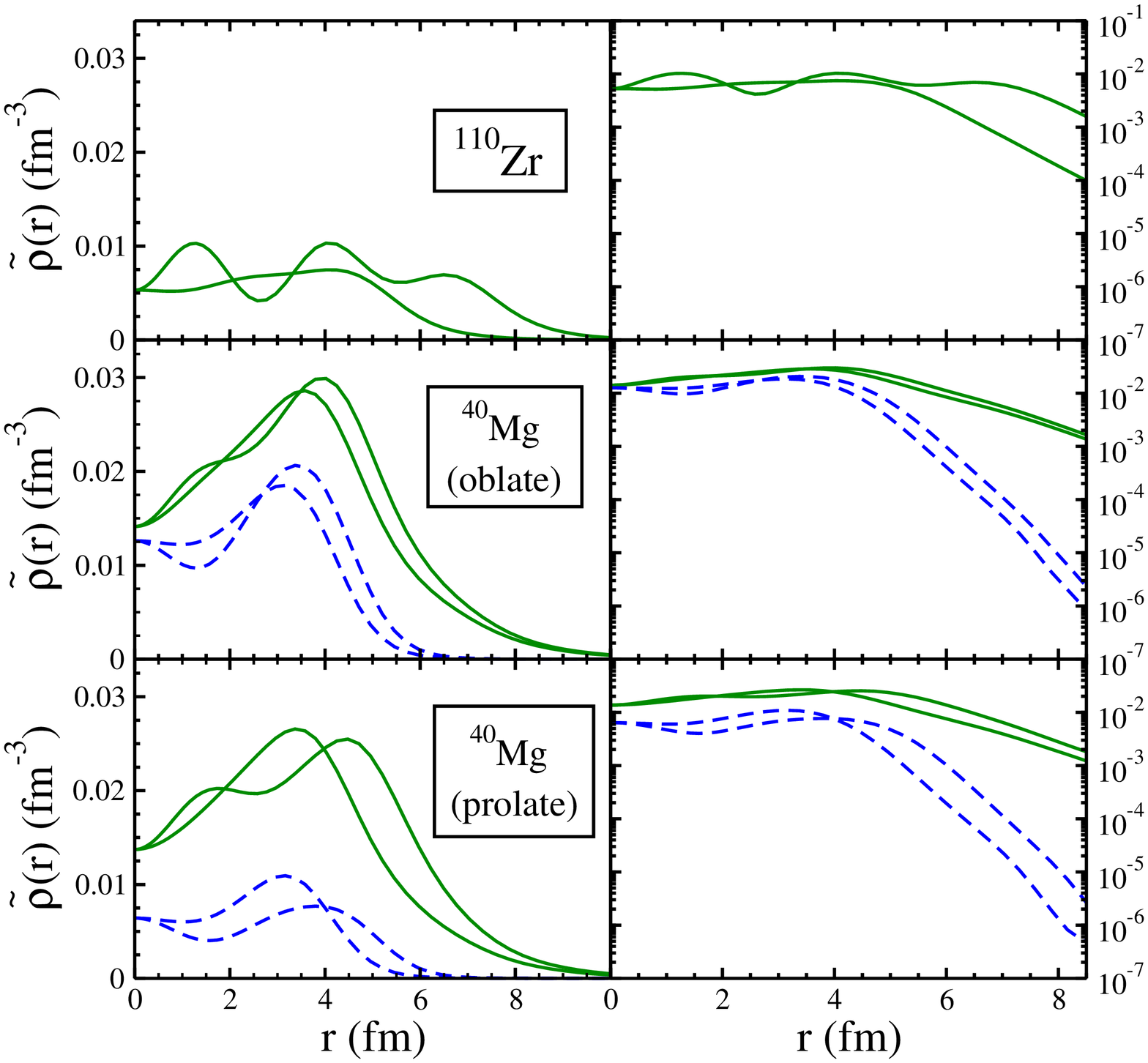}
\protect\caption{\label{fig_deformed_pair_dens}  (color online)
Same as in Fig.~\ref{fig_deformed_dens} but for pairing densities and without HFB/THO results. 
Proton pairing density is not represented for $^{110}$Zr as it is negligible therein.}
\end{figure*}

Results of the HFB/PTG calculation for a set of benchmark Ni isotopes close to the neutron drip line are presented in
Table~\ref{Ni_res_table}, Fig.~\ref{fig_spherical_densities} and Fig.~\ref{fig_spherical_pairing_densities}, 
where results of the HFB/Box calculation are also shown for comparison.
The Ni isotopes are spherical with  pairing in the neutron channel only. 
We see immediately a remarkable agreement between the results of the HFB/PTG and HFB/Box calculations.  
The difference in total energies is less than $85$ keV and the proton rms radii agree almost perfectly, 
while the neutron ones are slightly different by less than $0.003$~fm. Similarly good agreement is obtained for all other energy counterparts.
The good agreement in the ground state characteristics evaluated by the two different approaches is not surprising 
if one compares the density distributions shown in Fig.~\ref{fig_spherical_densities} and Fig.~\ref{fig_spherical_pairing_densities}. 
In these figures, the neutron and proton densities, $\rho_n$ and $\rho_p$, and the neutron pairing density $\tilde{\rho}_n$ are plotted
both in normal (left column) and logarithmic (right column)  scales. 
The agreement is almost perfect in the whole range of $r$ except at the box boundary where the HFB/Box densities vanish due to the boundary conditions
(however not seen in Fig~\ref{fig_spherical_pairing_densities}).
This agreement is striking considering the significant impact of the continuum for these nuclei and the fact that the HFB/PTG calculations 
are nevertheless performed using the basis expansion method.

Special attention has to be paid to the agreement for the pairing quantities. 
Interestingly, the pairing gap $\Delta_n$ increases as one approaches the drip line, indicating the important role of the pairing correlations in the continuum.
This result is somehow different from that of Ref.~\cite{giai} obtained by an alternative HFB calculation in the coordinate space 
for the same set of nuclei but it is in agreement with the estimates from Ref.~\cite{Yam05}.
In Fig.~\ref{fig_spherical_pairing_densities}, the scaling function of the THO basis is calculated with the method described in Ref.~\cite{sto04}, 
for which the quasi-exact density provided by the HFB/PTG calculations is used, and 16 THO shells are taken into account for each partial wave.
This implies virtually optimal results, and it has been checked that densities obtained from the HFB/Box and HFB/THO
methods are almost identical up to 20 fm. 
On the other hand, pairing densities given by the THO calculations are not 
exactly the same with those of the HFB/PTG and HFB/Box calculations,
as can be seen from Fig.~\ref{fig_spherical_pairing_densities}. 
While pairing densities  calculated with both methods for $^{84}$Ni and 
$^{90}$Ni are very close, those for $^{86}$Ni and $^{88}$Ni exhibit visible 
differences, especially for $^{86}$Ni, for which pairing energies differ 
by about 4 MeV.
Asymptotic properties of pairing densities calculated with the THO basis 
are also not well behaved after 15-20 fm, 
where they saturate instead of decreasing exponentially.
This indicates that THO basis calculations are not always devoid of 
inaccuracies, even at the spherical HFB level.

\subsection{Axially deformed nuclei}
In the case of axially deformed nuclei, few HFB/Box calculations are available to check the HFB/PTG results.
We consider the well-deformed nucleus $^{110}$Zr (deformation $\beta \approx 0.4$), 
already studied in Ref.~\cite{MarioUmar} and two states with different deformations for the drip line nucleus $^{40}$Mg. 
We use therein $k_{max}=$ 4 fm$^{-1}$ and $N_{\ell j} = 30$
for all partial waves.

Table~\ref{table5} compares the three approaches with respect to ground state properties of $^{110}$Zr. 
In general they yield similar values. The differences seen in Table~\ref{table5} are partially due to different structure of the model spaces adopted 
and the associated fitting of the pairing strength.

\begin{table}[ht]
\centering
\caption{Comparison of ground state properties of $^{110}$Zr calculated with the HFB/Box, HFB/PTG and HFB/THO approaches. 
The rms radii are in fm, quadrupole moments are in barn, and  all other quantities are in MeV.}
\label{table5}
\vspace{.2cm}
\tabcolsep=.2cm
\begin{ruledtabular}
\begin{tabular}{lrrrrrrrrrrrrrrr}
\noalign{\smallskip}
&  \multicolumn{1}{c}{HFB/Box}
&  \multicolumn{1}{c}{HFB/PTG}
&  \multicolumn{1}{c}{HFB/THO} \\
\hline \\
$Q_{tot}$     & 12.088	& 12.53	& 12.303 \\
$\Delta_n$    & 0.480   & 0.626 & 0.562         \\
$E^{pair}_n$  &   -1.53  & -3.015    & -2.05     \\
$r_n$   &  4.82	& 4.836 & 4.831              \\
$r_p$   &    4.55	& 4.560 & 4.556              \\
$E_{tot}$     &  -893.93 & -893.952 & -893.711  \\
\noalign{\smallskip}
\end{tabular}
\end{ruledtabular}
\end{table}

Proton and neutron densities for nuclei $^{110}$Zr and $^{40}$Mg are displayed 
in Fig.~\ref{fig_deformed_dens}, with comparison with THO results (circles)
for $^{110}$Zr, in normal scale (left column panels) and logarithmic scale 
(right column panels).
Associated pairing densities are shown in Fig.~\ref{fig_deformed_pair_dens}.

While agreement between the PTG and THO densities for $^{110}$Zr is good 
in normal scale, we can notice discrepancies in asymptotic properties, 
which are visible from the figure in logarithmic scale (see Fig.~\ref{fig_deformed_dens}). It is obvious that all densities calculated with the THO basis 
eventually follow the common asymptote dictated by the scaling function,
while they are well reproduced with use of the PTG basis.
This comparison also confirms the presence of deformation effects 
even in the far asymptotic region.

The middle and bottom panels in Figs.~\ref{fig_deformed_dens} and 
\ref{fig_deformed_pair_dens} illustrate the HFB/PTG normal and 
pairing densities 
for two states with different deformations in the drip line nucleus $^{40}$Mg. These states possess pairing correlations in both neutron and proton channels. 
The prolate and oblate states lead to asymptotic neutron densities 
which are very close,
as seen from the middle and bottom right panels in Fig.~\ref{fig_deformed_dens}.

\section{Conclusions}
\label{conclusion}
We have proposed a new method of the CHFB calculation for spherical and axially deformed nuclei, 
which properly takes the continuum into account. 
The method combines configuration-space diagonalization of the HFB Hamiltonian in the complete set of analytical 
PTG and Bessel/Coulomb wave functions with a matching procedure  in the coordinate space which restores the correct asymptotic 
properties of the HFB wave functions. 
The PTG potential is chosen to fit the nuclear HF potential and effective mass. The resulting PTG wave functions are close to
the bound and continuum states of the related HF potential while the resonance states are substituted by the bound 
PTG states with shifted single-particle energies. Partial waves of high angular momentum are very well represented by Bessel/Coulomb wave functions.

The main results of the present work are twofold:

First, we have obtained a new scheme (HFB/PTG) to solve the CHFB equations as a promising tool for large scale calculation; 
its performance is comparable, sometimes even better, to that of the HFB/THO code, for example. It properly takes the nuclear continuum into account 
and therefore could be used for precise density functional calculations for nuclei close to the drip lines. 
This HFB/PTG method can also be used to provide accurate quasiparticle wave
functions for microscopic calculations of dynamics beyond the nuclear mean-field approximation, as for example, the QRPA calculations for deformed nuclei.

Second, the fact that the HFB/PTG calculation reproduces the results of the coordinate-space HFB calculation with the box boundary condition (HFB/Box) 
even for nuclei up to the neutron drip lines is important. This result indicates the validity of the HFB/Box calculation which is widely used, 
although its validity is sometimes questioned when it is applied to the drip-line phenomena where continuum effects are crucially important \cite{giai}.

The inclusion of resonant structure in the basis is crucial for the success of the HFB/PTG approach. 
Our test calculations indicate significant disagreement with the HFB/Box result 
if the PTG bound states representing the resonant GHF states are removed from the basis: 
in their absence, the pairing densities are overestimated in the surface region, while particle densities are slightly underestimated in the inner region. 
This means that the resonance states significantly contribute to the total energy through both the particle-hole and particle-particle channels. 
Their contributions to the pairing correlation energy are evaluated to be about 2-3 MeV for the case of Ni isotopes close to the neutron drip line.

A more complete investigation of the importance of the HFB resonance states could be made by a detailed comparison 
with the result of the exact Gamow-HFB calculation. Such an analysis is in progress for spherical nuclei and will be reported in the near future \cite{nicolas} .

\section*{Acknowledgments}
The authors acknowledge Japan Society for the Promotion of Science for awarding The Invitation Fellowship for Research in Japan (Long-term) to M.~S. 
and The JSPS Postdoctoral Fellowship for Foreign Researchers to N.~M., which make our collaboration possible. 
This work was supported by the JSPS Core-to-Core Program ``International 
Research Network for Exotic Femto Systems." 
This work was carried out as a part of the U.S.~Department of Energy under Contracts Nos.~DE-FG02-96ER40963 (University of Tennessee),
DE-AC05-00OR22725 with UT-Battelle, LLC (Oak Ridge National Laboratory), and DE-FG05-87ER40361 (Joint Institute for Heavy Ion Research), 
the UNEDF SciDAC Collaboration supported by the U.S. Department of Energy under grant No. DE-FC02-07ER41457.

\appendix{}
\section{PTG basis}
\label{PTG_appendix}
\subsection{PTG potential}
The one-body Hamiltonian for the exactly solvable PTG model reads:
\begin{eqnarray}
H_{PTG} &=& \frac{\hbar^2}{2 m_0} \left( - \frac{d}{dr} \frac{1}{\mu(r)}
\frac{d}{dr} + \frac{\ell(\ell+1)}{r^2 \mu(r)} \right) \nonumber \\
&+& V_{PTG}(r)
\label{ptgham}
\end{eqnarray}
with $m_0$ the particle free mass, $r$ is the radial coordinate (in fm),
$\mu(r)$ its dimensionless effective mass
(the full effective mass is $m_0 \; \mu(r)$),
$\ell$ its orbital angular momentum and $V_{PTG}$ is the PTG potential.
The potential $V_{PTG}(r)$ and the effective mass $\mu(r)$ are written:
\begin{eqnarray}
\mu(r) &=& 1 - a(1-y^2) , \label{mu_r} \\
V_{PTG}(r) &=& \frac{\hbar^2 s^2}{2 m_0 \mu(r)} \nonumber \\
&\times& \left( V_{\mu}(r) + V_{\ell}(r) + V_{c}(r) \right),
\label{V_PTG}
\end{eqnarray}
where 
$s$ is the scaling parameter, 
$V_{\mu}$ the potential part issued from the effective mass,
$V_{\ell}$ its $\ell$-dependent part and
$V_{c}$ its main central part, defined by
\begin{eqnarray}
V_{\mu}(r) &=& \left[ 1-a + \left[ a(4-3 \Lambda^2) \right. \right. \nonumber \\
&-& \left. 3(2 - \Lambda^2) \right]  y^2 \nonumber \\
&-& \left. (\Lambda^2 - 1)(5(1-a) + 2 a y^2) \; y^4 \right] \nonumber \\
&\times& \frac{a}{\mu(r)^2} (1-y^2) \left[ 1 + (\Lambda^2 - 1) y^2 \right] , \\
V_{\ell}(r) &=& \ell(\ell+1)
\left[ \frac{(1 - y^2)( 1 + (\Lambda^2 - 1) y^2)}{y^2} \right. \nonumber \\
&-& \left. \frac{1}{s^2 r^2} \right]
\mbox{ , } r > 0 \label{Vl_r} , \\
V_{c}(r) &=& (1-y^2) \left[ -\Lambda^2 \nu (\nu+1) \right. \nonumber \\
&-& \frac{\Lambda^2 - 1}{4} \left( 2 - (7 - \Lambda^2) y^2  \right. \nonumber \\
&-& \left. \left. 5(\Lambda^2 - 1) y^4 \right)  \right] .
\end{eqnarray}
The quantities $V_{PTG}(r)$ and $\mu(r)$ depend on
an implicit function $y=y(r)$ defined in the following way:
\begin{equation}
\begin{array}{lll}
\displaystyle \Lambda^2 s~ r &=& \mbox{arctanh}(y) + \sqrt{\Lambda^2 - 1}
\mbox{ } \arctan(\sqrt{\Lambda^2 - 1} \; y)
\label{ptgy}
\end{array}
\end{equation}
so that ${0 \leq y < 1}$ for ${0 \leq r< \infty}$.

The numerical solution of Eq.~(\ref{ptgy})
by way of Newton/bisection methods is stable but one should take
special care at large distances when $y$ becomes closely equal to one.
For example, this can be done by rewriting  Eq.~(\ref{ptgy}),
introducing the new variable $x = \mbox{arctanh}(y)$:
\begin{equation}
\begin{array}{lll}
\displaystyle \Lambda^2 s~ r&=& x \\
&+&\sqrt{\Lambda^2 - 1} \mbox{ } \arctan(\sqrt{\Lambda^2 - 1} \tanh(x)),
\label{ptgxy}
\end{array}
\end{equation}
It is solved with respect to $x$ with a fixed-point algorithm.  In this region,
$1-y^2$ should be calculated in terms of the expression
${1-y^2=4 e^{-2x}/(1+e^{-2x})^2}$ to avoid numerical cancellations.

One has to mention that, in the calculation of $V_{PTG}(r)$,
$V_{\ell}(r)$ is finite for all $r \geq 0$ but is the difference of two
diverging terms for $r \rightarrow 0$.
Thus, to be precise in this region, Eq.~(\ref{ptgy}) must be rewritten
as a power series in $y$, so that
the main diverging terms of Eq.~(\ref{Vl_r}) cancel analytically.

As seen from the equations above, there are four parameters in the PTG model:
the effective mass parameter $a$, the scaling parameter $s$, the parameter
$\Lambda$ determining the shape of the potential and the parameter $\nu$
associated with the depth of the potential.
They can take different values for different angular momenta $\ell$.
We can use this freedom in order to approximate the nuclear
potential for each $\ell j$-subspace,
as described in Sec.~\ref{HFB_PTG_scheme_section}.

\subsection{PTG states}
The PTG wave functions and eigen-energies are determined by
the Schr\"odinger equation for the Hamiltonian  (\ref{ptgham})
\begin{equation}
\displaystyle H_{PTG}~ \Psi_k(r) = E~\Psi_k(r)
\label{pteq}
\end{equation}
with energies
\begin{equation}
 \displaystyle
E=\frac{\hbar^2 k^2 }{2m_0},
\label{ptgenergy}
\end{equation}
where $k$ stands for the complex linear momentum associated with $E$.

For bound states, if they exist, the parameter $\nu$ determines the maximal
value $n_{max}$ of the radial quantum number $n=0,1,2,...,n_{max}$
as the largest integer inferior to
\begin{equation}
\displaystyle
\left\{\frac{1}{2} \left( \nu-\ell-\frac{3}{2} \right) \right\},
\label{ptgnmax}
\end{equation}
and defines the complex momentum
\begin{equation}
 \displaystyle
k_{nl} = i s \frac{- A_{nl} + \sqrt{\Delta_{nl}}}{1-a},
\label{ptgkb}
\end{equation}
with
\begin{eqnarray}
\displaystyle A_{nl} &=& 2n+\ell+\frac{3}{2}, \\
\Delta_{nl} &=& \Lambda^2 \left( \nu + \frac{1}{2} \right)^2 (1-a) \nonumber \\
&-& \left[ (1-a)\Lambda^2 - 1 \right] A_{nl}^2 .
\end{eqnarray}
For continuum states, $k$ can take any
real positive values from zero to infinity.


\subsection{PTG wave functions}
In order to express the PTG wave function  $\Phi_k(r)= r~ \Psi_k(r)$
in a closed analytical form, let us introduce the following three functions
\begin{equation}
f_k(r)=  \displaystyle
F\left(\nu^-,\nu^+,\ell
+\frac{3}{2},x^- \right)\left( x^+ \right)^{\bar{\beta}/2},
\label{ptgf}
\end{equation}
\begin{equation}
f_k^+(r)=  \displaystyle
F\left(\nu^-,\nu^+,\bar{\beta}+1,x^+
\right)\left( x^+ \right)^{\bar{\beta}/2},
\label{ptgfp}
\end{equation}
\begin{equation}
f_k^-(r)=  \displaystyle
F\left(\mu^-,\mu^+,-\bar{\beta}+1,x^+ \right)\left( x^+ \right)^{-\bar{\beta}/2}
\label{ptgfm}
\end{equation}
and
\begin{equation}
\displaystyle \chi_k(r) = \sqrt{\frac{x^- + \Lambda^2(1-a)x^+}{\sqrt{x^-
+ \Lambda^2 x^+}}} (x^-)^{\frac{\ell+\frac{3}{2}}{2}},
\label{ptgchi}
\end{equation}
where
\begin{equation}
\begin{array}{lll}
\displaystyle x=\frac{1-(\Lambda^2+1)y^2}{1+(\Lambda^2-1)y^2}, x^-
= \frac{1-x}{2}, x^+ = \frac{1+x}{2},
\label{ptgx}
\end{array}
\end{equation}
\begin{equation}
 \displaystyle
\nu^+=\frac{\ell+\frac{3}{2}+\bar{\beta}+\bar{\nu}}{2},~~
\nu^-=\frac{\ell+\frac{3}{2}+\bar{\beta}-\bar{\nu}}{2},
\end{equation}
\begin{equation}
\displaystyle
\mu^+=\frac{\ell+\frac{3}{2}-\bar{\beta}+\bar{\nu}}{2},~~
\mu^-=\frac{\ell+\frac{3}{2}-\bar{\beta}-\bar{\nu}}{2},
\end{equation}
\begin{equation}
 \displaystyle
\bar{\beta}=-\frac{i k}{\Lambda^2 s},
\label{ptgbeta}
\end{equation}
\begin{equation}
 \displaystyle
\bar{\nu}=\sqrt{(\nu+1/2)^2+\bar{\beta}^2(1-\Lambda^2(1-a))},
\label{ptgnu}
\end{equation}
and $F(a,b,c,z)$ is the Gauss hypergeometric function \cite{Abr70}.

In the case of bound states, $k_{nl}$ determines the momenta $k$
which are pure imaginary (see Eq.~(\ref{ptgkb})),
while they are real positive numbers in the case of scattering states.
This defines all other quantities entering the equations above.
For both cases, the PTG wave functions can be written either as
\begin{equation}
\displaystyle
\Phi_k(r)= \mathcal{N} ~\chi_k(r)~ f_k(r)
\label{ptgwf1}
\end{equation}
or as
\begin{equation}
\displaystyle
\Phi_k(r)=  \mathcal{N} \chi_k(r)~ \left( A^+ ~ f^+_k(r) + A^- ~ f^-_k(r)
\right).
\label{ptgwf2}
\end{equation}
Equation (\ref{ptgwf1}) is suitable for numerical work for small distances
since $x^- \rightarrow 0$  when  $r\rightarrow 0$
so that one is away from the pole
of the hypergeometric function appearing at $x^- = 1$.
Similarly, Eq.~({\ref{ptgwf2}}) is applicable for large
distances since  $x^+ \rightarrow 0$ when $r \rightarrow +\infty$ and
the pole $x^+ = 1$ of the hypergeometric function in
Eqs.~(\ref{ptgfp}) and (\ref{ptgfm}) is avoided.

In the case of bound states, the quantum numbers $\{n \ell\}$ are the principal
quantum number $n$ and the angular momentum $\ell$.
The constants ${\mathcal{N},A^+,A^-}$
entering Eqs.~(\ref{ptgwf1}) and (\ref{ptgwf2}) are given by:
\begin{eqnarray}
\mathcal{N} &=& \sqrt{\frac{2 \Lambda^2 s \bar{\beta} (\ell
+ \frac{3}{2} + \bar{\beta} + 2n)}
{(\ell + \frac{3}{2} + \bar{\beta} \Lambda^2 (1-a) + 2n)}}  \nonumber \\
&\times& \sqrt{\frac{\Gamma(\ell + \frac{3}{2} + \bar{\beta} + n)
\Gamma(\ell + \frac{3}{2} + n)}
{\Gamma(n+1) \Gamma(\bar{\beta} + n + 1) \Gamma(\ell + \frac{3}{2})^2}} ,
\nonumber \\
A^+ &=& \frac{\Gamma(\ell + \frac{3}{2}) \Gamma(- \bar{\beta})}
{\Gamma(\mu^+) \Gamma(\mu^-)}, A^- = 0 ,
\label{ptg_const_bound}
\end{eqnarray}
where $\Gamma(z)$ is the Gamma function \cite{Abr70}.

In the case of scattering states, the quantum numbers $\{k \ell\}$
include the momentum $k$ and the angular momentum $\ell$
while the associated constants  ${\mathcal{N},A^+,A^-}$ read:
\begin{eqnarray}
&& \mathcal{N} = \sqrt{\frac{\Gamma(\nu^+) \Gamma(\nu^-) \Gamma(\mu^+)
\Gamma(\mu^-)}
{2 \pi \; \Gamma(\bar{\beta}) \Gamma(- \bar{\beta})
\Gamma(\ell + \frac{3}{2})^2}} \nonumber \\
&&A^+ = \frac{\Gamma(\ell + \frac{3}{2}) \Gamma(- \bar{\beta})}
{\Gamma(\mu^+) \Gamma(\mu^-)} \nonumber \\
&&A^- = \frac{\Gamma(\ell + \frac{3}{2}) \Gamma(\bar{\beta})}
{\Gamma(\nu^+) \Gamma(\nu^-)}.
\label{ptg_const_scat}
\end{eqnarray}

The normalization constant $\mathcal{N}$ is determined from the
normalization condition
\begin{equation}
\displaystyle
\int_0^\infty \Phi_{nl}(r)\Phi_{n'l}(r) dr = \delta_{nn'}
\label{ptgBboundnorm}
\end{equation}
for bound states and from the Dirac delta function normalization
for scattering states:
\begin{equation}
\displaystyle
\int_0^\infty \Phi_{kl}(r)\Phi_{k'l}(r) dr = \delta(k-k')
\label{ptgdeltanorm}
\end{equation}

All bound and scattering wave functions are orthogonal to each other
\begin{equation}
\displaystyle
\int_0^\infty \Phi_{k}(r)\Phi_{k'}(r) dr = 0,~~~k\neq k'
\label{ptgorthogonality}
\end{equation}
and they form a complete basis
\begin{equation}
\begin{array}{lll}
&\displaystyle \sum_{nl} \Phi_{nl}(r)\Phi_{nl}(r')  \\
+&\displaystyle \sum_{l}\int_0^\infty \Phi_{kl}(r)\Phi_{kl}(r') dk
= \delta(r-r').
\end{array}
\label{ptgBcomplete}
\end{equation}

One can check that at large distances
\begin{equation}
\displaystyle
x\rightarrow -1+ 2 e^{-2\Lambda^2 s (r-r_1)}, ~~~ r \rightarrow +\infty ,
\label{ptgBxass}
\end{equation}
where
\begin{equation}
\displaystyle
\Lambda^2 s~r_1=\sqrt{\Lambda^2-1} \arctan (\sqrt{\Lambda^2-1})
-\log \left( \frac{\Lambda}{2} \right).
\label{ptgr1}
\end{equation}
Substituting this into Eq.~({\ref{ptgwf2}}) one obtains the asymptotic
form of the PTG wave functions
\begin{equation}
\displaystyle
\Phi_k(r)\mapsto C^+ ~ e^{i k r} ~+~ C^- ~ e^{- i k r}
\label{ptgwfass}
\end{equation}
where
$C^+ = \mathcal{N} A^+ e^{-i k r_1}$ and $C^- = \mathcal{N} A^- e^{i k r_1}$,
defined by Eqs.~(\ref{ptg_const_bound}) and (\ref{ptg_const_scat}).

The PTG wave functions are numerically stable and accurate when using
Eq.~(\ref{ptgwf1}) up to $y\leq 0.99$ then applying the form (\ref{ptgwf2}).
They accurately land onto their asymptotic representation
of Eq.~(\ref{ptgwfass}) at large distances.

\section{Matrix elements}
\label{ME_appendix}
Let us deal with numerical integration in $r$ and $k$ space.
The integration in the $r$ space is performed in terms of $N_{r}$
Gauss-Legendre integration points $x_i$ and weights $w_i$ within
the interval $[0,R_{max}]$,
\begin{eqnarray}
\displaystyle
&&\int_0^\infty ~O(r)~\Phi_{k}(r)~\Phi_{k'}(r)~dr \nonumber \\
&\simeq& \sum_{i=1}^{N_{r}} ~O(r_i)~\Phi_{k}(r_i)~\Phi_{k'}(r_i) w_i,
\end{eqnarray}
where $O(r)$ is an arbitrary function of $r$ and $R_{max}$ is a point where
nuclear potential disappears. Usually a value $R_{max}=15$ fm is used.
In the same way, integration in the $k$ space is done in terms of $N_{k}$
Gauss-Legendre integration points $k_i$ and weights $w_{k_i}$
within the interval $[0,k_{max}]$,
\begin{eqnarray}
\displaystyle
&&\int_0^{k_{max}} ~O(k)~\Phi_k(r)~\Phi_k(r')~dk  \nonumber \\
&\simeq& \sum_{i=1}^{N_{k}} ~O(k_i)~\Phi_{k_i}(r)~\Phi_{k_i}(r')~w_{k_i},
\end{eqnarray}
where $O(k)$ is an arbitrary function of $k$.

Radial integrals must be calculated cautiously due to the presence of
non-integrable scattering states in the basis.
When the radial operator represents the nuclear potential
or explicitly depends on nuclear densities or currents,
one can safely integrate the matrix
elements to some large but finite distance $R_{max}$. Beyond $R_{max}$,
the contribution of the integral becomes negligible due to the presence of
the densities or currents. However, it is not the case
for the kinetic + Coulomb part of the Hamiltonian.
This operator is infinite-ranged and
induces Dirac delta functions in the matrix elements,
which have to be regularized
directly. For this, one separates the matrix element in two integrals,
defined on the intervals
$[0:R_{max}]$ and $[R_{max}:+\infty[$. The first part is finite
and treated with standard methods.
For the second part, if one deals with Bessel/Coulomb wave functions,
one can assume that the nuclear part is negligible
after $R_{max}$ so that they are solutions
of the asymptotic HF equations. Hence, one obtains:
\begin{eqnarray}
&&\int_{R_{max}}^{+\infty} u_\alpha(r) h(r) u_\beta(r) \;
dr \nonumber \\
&=& k_\alpha^2 \left( \delta_{\alpha \beta}
- \int_0^{R_{max}} \!\!\!\!\! u_\alpha(r) u_\beta(r) \;
dr \right) \mbox{ (bound)}  \nonumber \\
&=& k_\alpha^2 \left( \delta(k_\alpha - k_\beta)
- \int_0^{R_{max}} \!\!\!\!\! u_\alpha(r) u_\beta(r) \; dr \right)
\mbox{ (scat)}  \nonumber \\
&=& -k_\alpha^2 \int_0^{R_{max}} \!\!\!\!\! u_\alpha(r) u_\beta(r) \;
dr \mbox{ (mixed)} \label{asymp_int}
\end{eqnarray}
where $h(r)$ is the HF potential
which reduces to the kinetic + Coulomb Hamiltonian asymptotically.
Here, ``bound'' (``scat'') means that both $\alpha$ and $\beta$ states are
bound (scattering) and ``mixed'' means that
$\alpha$ is bound and $\beta$ scattering or vice-versa.
The Dirac delta with a discretized basis becomes
$\delta_{\alpha \beta}/w_{k_\alpha}$
with $w_{k_\alpha}$ being the Gauss-Legendre weight
associated to the discretized value $k_\alpha$,
so that its implementation is immediate;
since all integrals are finite, they pose no problem.
When the PTG basis states are used
instead of the Bessel/Coulomb wave functions,
it turned out that it is numerically precise to disregard
the Coulomb/centrifugal part of the Hamiltonian after $R_{max}$,
so that Eq.~(\ref{asymp_int}) is the same for both
the PTG and Bessel/Coulomb wave functions.
Indeed, Eqs.~(\ref{ptgBxass}) and (\ref{ptgwfass}) imply that
the PTG wave functions behave asymptotically like neutron waves functions
of angular momentum $\ell = 0$.
The above seemingly crude approximation can, in fact, be mathematically
justified. The HFB matrix evaluated using such a procedure
converges weakly to the exact HFB matrix
for $R_{max} \rightarrow +\infty$ \cite{weak_cv_book}.
This means that the HFB matrix elements depend on $R_{max}$ asymptotically,
some of them even diverging with $R_{max} \rightarrow +\infty$,
whereas its eigenvalues and eigenvectors converge to a finite value.


\end{document}